\documentclass[twocolumn,secnumarabic,amssymb, nobibnotes, 
aps, 
longbibliography
]{revtex4-2}

\usepackage{graphicx,footmisc}
\usepackage[caption=false]{subfig}
\usepackage{color,mwe}
\usepackage{natbib,url}
\usepackage{amsmath}
\usepackage{float}
\usepackage[normalem]{ulem}

\newcommand*{\be}{\begin{equation}}
\newcommand*{\ee}{\end{equation}}

\def\begineq{\begin{equation}}
\def\endeq{\end{equation}}

\def\begineqn{\begin{equation*}}
\def\endeqn{\end{equation*}}

\def\beginar{\begin{eqnarray}}
\def\endar{\end{eqnarray}}
\def\beginarn{\begin{eqnarray*}}
\def\endarn{\end{eqnarray*}}

\def\lb{\left ( }
\def\rb{\right ) }

\def\Rat{\widetilde{Ra}}
\def\Rmt{\widetilde{Rm}}

\def\ub{\mathbf{u}}
\def\Zb{\boldsymbol{\zeta}}

\def\He{\overline{\mathcal{H}}}
\def\Bb{\mathbf{B}}
\def\ubp{\mathbf{u}^{\prime}}

\def\mub{\overline{\bf u}}
\def\mBb{\overline{\bf B}}
\def\pBb{\mathbf{B}^{\prime}}

\def\hz{{\bf\widehat z}}

\def\Ret{\widetilde{Re}}

\begin{document}

\title{Strong large scale magnetic fields in rotating convection-driven dynamos: the important role of magnetic diffusion}

\author{Ming Yan and
Michael A. Calkins
}

\affiliation{
Department of Physics, University of Colorado, Boulder, CO  80309, USA \\
}

\begin{abstract}
Natural dynamos such as planets and stars generate global scale magnetic field despite the inferred presence of small scale turbulence. Such systems are known as large scale dynamos and are typically driven by convection and influenced by rotation. Previous numerical studies of rotating dynamos generally find that the large scale magnetic field becomes weaker as the flow becomes more turbulent. The underlying physical processes necessary for sustaining so-called large scale dynamos is therefore still debated. Here we use a suite of numerical simulations to show that strong large scale magnetic fields can be generated in rotating convective turbulence provided that two conditions are satisfied: (1) the flow remains rotationally constrained; and (2) magnetic diffusion is important on the small convective length scale. These findings are in agreement with previous asymptotic predictions and suggest that natural dynamos might satisfy these two conditions.
\end{abstract}

\maketitle

Planets and stars generate magnetic field through the dynamo process that converts the kinetic energy of fluid motion into magnetic energy \citep[e.g.][]{cJ11b,sT21}. These fields are often dominated by their global (e.g.~dipolar) components, whereas the underlying fluid motion that powers the dynamo is turbulent and characterized by length scales much smaller than the global scale. The Coriolis force likely plays a fundamental role in the generation of such large scale structure by leading to domain-scale correlations in the flow field that generate coherent large scale magnetic field \citep{eP55,mS66a,hM72}. This correlation can be quantified by the helicity of the fluid motion \citep{hM70}, defined as the dot product between the velocity field and the curl of the velocity field. However, simulations tend to find that the relative contribution of the large scale component of the dynamo generated magnetic field decreases as the fluid becomes more strongly forced and turbulent \citep{uC06,fC06,aT12,bF13b}. This loss of a predominantly large scale magnetic field is associated with a decrease in the relative (as measured to a maximum value) helicity. The fundamental question pertaining to natural systems is then how a large scale dynamo is maintained in the presence of strongly forced, small scale turbulence. Asymptotic theory relevant to rapidly rotating flows suggests that the deficit in relative helicity can be overcome by enhancing the influence of magnetic diffusion on the small convective length scale \citep{mC15b}. In the present work we utilize direct numerical simulations (DNS) of rapidly rotating convective turbulence to confirm these asymptotic predictions, thus helping to shed light on a major problem in dynamo theory.

Buoyancy is thought to provide the main source of power for natural dynamos. Rayleigh-B\'enard convection, consisting of a fluid layer of depth $H$ heated from below and cooled from above, is a canonical system for studying buoyancy-driven turbulence and dynamos. The buoyancy force is controlled by the non-dimensional 
Rayleigh number,
\be
Ra = \frac{g \alpha \Delta T H^3}{ \nu \kappa },
\ee
where $g$ is the (constant) gravitational acceleration, $\alpha$ is the thermal expansion coefficient, $\Delta T$ is the temperature difference between the top and bottom boundaries, $\nu$ is the kinematic viscosity and $\kappa$ is the thermal diffusivity. The relative importance of viscous and inertial forces to the Coriolis force is quantified by the Ekman number and the Rossby number defined by, respectively,
\be
E = \frac{\nu}{2 \Omega H^2}, \quad Ro = \frac{U}{2 \Omega H},
\ee
where $\Omega$ is the system rotation rate and $U$ is the characteristic dimensional flow speed. The Reynolds number, quantifying the relative importance of inertia and viscous forces, is then related via $Re = Ro/E$. 

The electrical properties of the fluid are specified by the magnetic Prandtl number, $Pm = \nu/\eta$, where $\eta$ is the magnetic diffusivity. Dynamos require sufficiently large flow speeds to overcome, or at least balance, the resistive effects of ohmic diffusion. Therefore, the magnetic Reynolds number, $Rm = Re Pm$, must exceed a threshold value so that the self-induced magnetic field does not diffuse away. Planetary and stellar dynamos are in a regime of rapidly rotating magnetized convective turbulence in which $E \ll Ro \ll 1$ and $Re \gg Rm \gg 1$, which is a parameter space that is challenging to study numerically given the broad range of scales characterizing the dynamics. 

Asymptotic theory can provide insight into the extreme parameter space that characterizes most natural dynamos. A brief summary of this theory is provided here to aid in the interpretation of the simulation results. For details the reader is referred to Refs.~\citep{mC15b,mC17b}. In the limit $E \rightarrow 0$ and $Ro \rightarrow 0$, 
the dynamics of convection depend on the reduced Rayleigh number \citep[e.g.][]{kJ12}
\be
\Rat \equiv Ra E^{4/3}.
\ee
At leading asymptotic order the flow is geostrophically balanced on $O\lb E^{1/3} \rb$ horizontal convective length scales. Thus, when the governing equations are non-dimensionalized on this small convective length scale, horizontal derivatives are $O\lb 1 \rb$ and vertical derivatives are $O \lb E^{1/3} \rb$. Variables are decomposed into horizontal averages (mean) and fluctuating quantities such that the magnetic field and velocity field become $\Bb = \mBb(z,t) + \pBb(x,y,z,t)$ and $\ub = \mub(z,t) + \ubp(x,y,z,t)$, respectively. In the plane layer geometry studied here the mean velocity field is negligibly small and so $\ub \approx \ubp$. In the asymptotic regime, the small scale magnetic Reynolds number, defined as 
\be
\Rmt = Rm E^{1/3},
\ee
plays the key role in determining the relative sizes of the mean and fluctuating magnetic field. In particular, accessing what we refer to as the energetically robust large scale dynamo regime, in which the energy contained in the large scale (mean) component of the magnetic field is asymptotically larger than the energy contained in the fluctuating magnetic field, is achieved by balancing the emf with large scale magnetic diffusion, $\hz \times \partial_z \overline{\lb \ubp \times \pBb \rb} \sim \Rmt^{-1} \partial_z^2 \mBb$, and mean stretching with small scale magnetic diffusion, $\mBb \cdot \nabla \ub \sim \Rmt^{-1} \nabla^2 \pBb $; these two balances give, respectively,
\be
|\pBb| / |\mBb| \sim E^{1/3}/\Rmt, \qquad |\pBb| / |\mBb| \sim \Rmt ,
\ee
where we have used the fact that $|\ubp| = O(1)$ to ensure geostrophic balance at leading order \citep{mC15b}. For the two relations given above to be consistent we then require
\be
\Rmt = O\lb E^{1/6} \rb .
\label{E:Rmt}
\ee
This relationship states that the energetically robust large scale dynamo regime requires that both $E$ and $\Rmt$ are small, i.e.~equation \eqref{E:Rmt} is the distinguished limit that allows for the large scale magnetic field to be energetically larger than the small scale magnetic field. 

Connecting the asymptotic scaling of equation \eqref{E:Rmt} with $Pm$ can be made upon noting that $\Rmt = Pm \Ret$, where $\Ret  = Re E^{1/3}$ is the small scale Reynolds number. In a given simulation, $\Ret$ is controlled by $\Rat$. Thus, reaching small values of $\Rmt$ requires that $Pm$ is reduced; since $\Ret = O(1)$ this implies we need $Pm = O\lb E^{1/6} \rb$. Importantly, the energetically robust large scale dynamo regime is not limited to small values of $\Rat$ (i.e.~limited to near the onset of convection), and is therefore not limited to convective states in which the helicity is maximal. We note that the simulations are consistent with a $E^{1/3}$ scaling for the convective length scale, though a dependence on $\Rat$ is also observed. The inertial theory for rotating convection suggests that the length scale should behave like $Ro^{1/2}$ \citep[e.g.][]{cG19,jmA20}. Though we do not test this inertial scaling with the present simulations, such a scaling would still imply that $\Rmt$ must be small to observe strong large scale magnetic field.

In the present study we simulate dynamo action of a Boussinesq fluid driven by rotating convection as $E $, $Ra$, and $Pm$ are varied. Reduced Rayleigh numbers up to $\Rat \approx 80$ are reached for each value of $E$; the so-called geostrophic turbulence regime occurs for $\Rat \gtrsim 40$ \citep{kJ12}. For Ekman numbers within the range $10^{-6} \le E \le 10^{-4}$ we use $Pm=1$, and for $E \le 10^{-6}$ we reduce the magnetic Prandtl number down to $Pm=0.05$ at the smallest Ekman number considered, $E =10^{-8}$. The largest Reynolds number achieved in the survey was $Re = 1.53 \times 10^4$ for $E =10^{-8}$ and $\Rat \approx 80$. The maximum Rossby was $Ro = 0.06$ for $E =10^{-4}$ and $\Rat \approx 80$. The simulations use a de-aliased pseudospectral method in which all flow variables are expanded as Chebyshev polynomials in the vertical dimension and Fourier series in the horizontal dimensions. Resolutions up to $432$ Fourier modes and $864$ Chebyshev polynomials were used for the most demanding calculations. A third order accurate implicit-explicit time-stepping scheme is used \citep[e.g.][]{pM16,mY19} in which timestep sizes (in non-dimensional rotation time) down to $\Delta t = 5\times10^{-4}$ were used. Each simulation was run until a statistically stationary state was achieved and statistics were computed only once this state was reached. We found that this state was reached in approximately hundreds to thousands of rotation times. Approximately 5 million CPU hours were required to generate the dataset used in the present study. In all cases shown the horizontal dimension of the simulation domain is scaled such that 10 critical wavelengths are present. The boundary conditions are stress-free, isothermal and the magnetic field is required to be purely vertical at the top and bottom boundaries. The thermal Prandtl number is fixed at $Pr = \nu / \kappa =1$.

\begin{figure*}
 \begin{center}
       \subfloat[]{
      \includegraphics[width=0.45\textwidth]{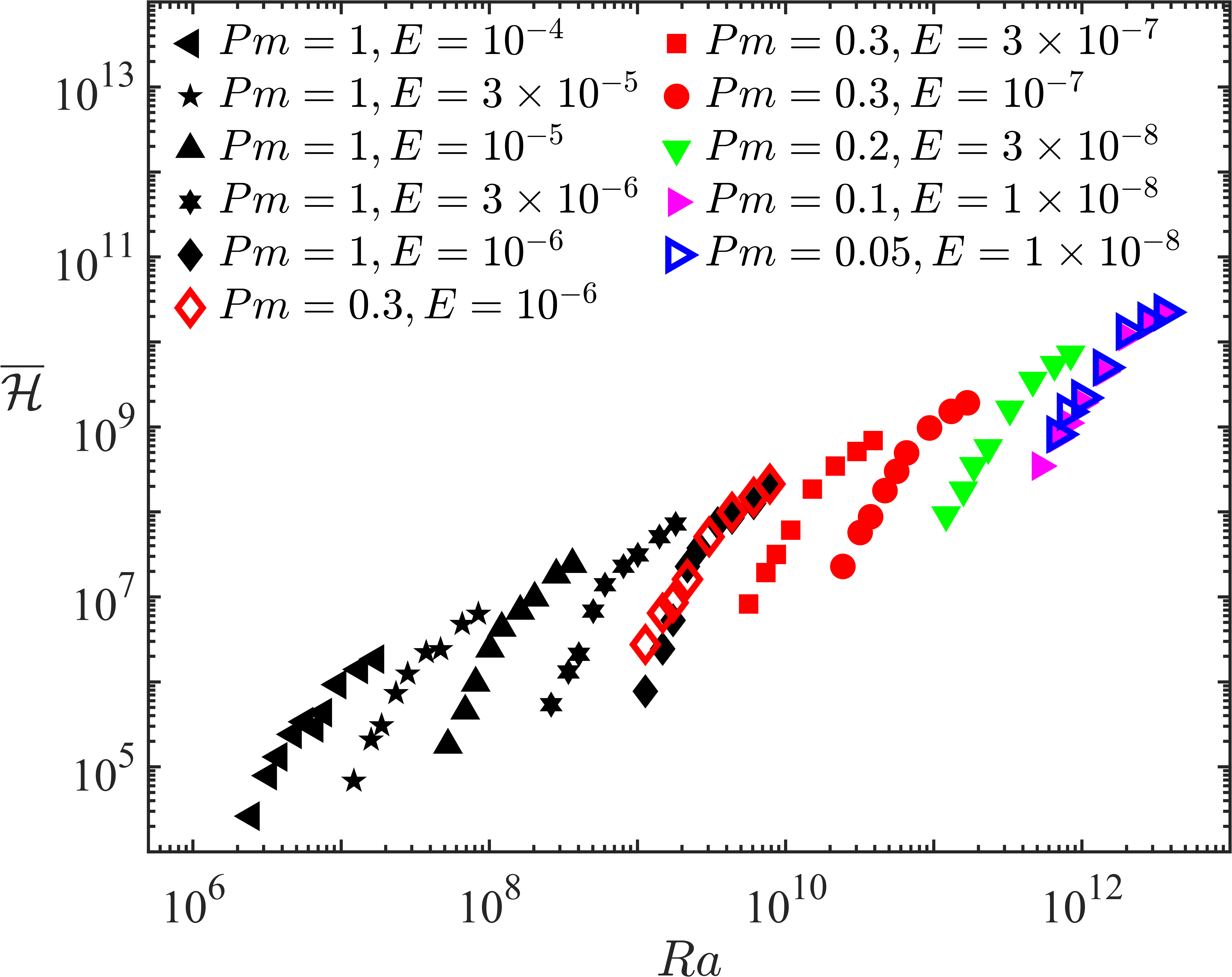}} \qquad
       \subfloat[]{
      \includegraphics[width=0.46\textwidth]{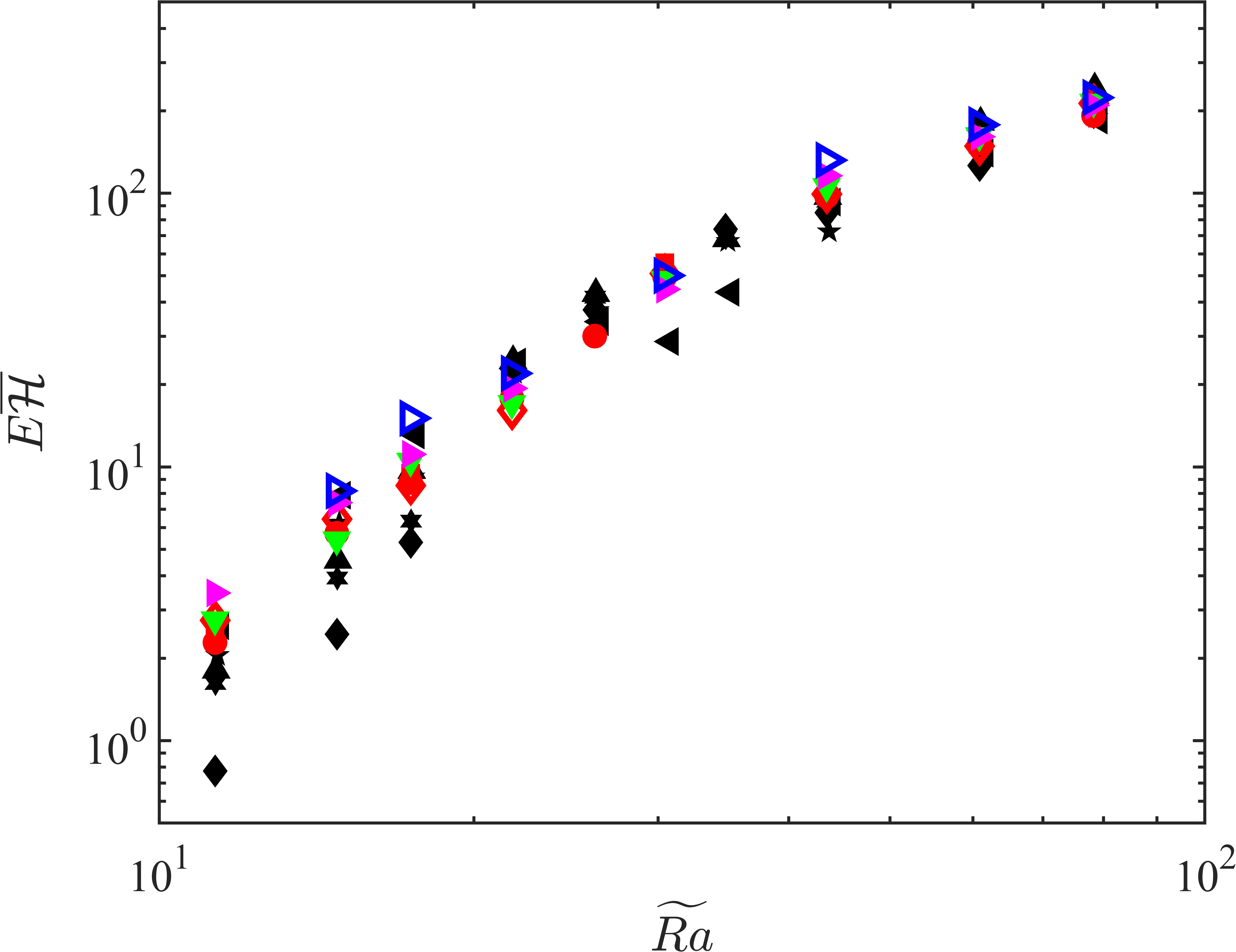}}
 \end{center}
\caption{ Helicity (rms) for all simulations: (a) $\He$ versus $Ra$; (b) asymptotically rescaled rms helicity $E \He $ versus reduced Rayleigh number, $\Rat = Ra E^{4/3}$. For the symbols the colors denote different values of $Pm$ and the shapes denote different values of $E$.
\label{F:He}}
\end{figure*} 


The mean helicity is denoted by $\He (z) \equiv \overline{\ubp \cdot \Zb'}$, where $\Zb' = \nabla \times \ub'$ is the vorticity. Helicity can stretch and twist magnetic field lines collectively over the horizontal plane, and therefore induce large scale magnetic field. It is well known that when strongly influenced by rotation, convection is helical \citep[e.g.][]{tV21}. Fig. \ref{F:He}(a) shows the rms value of the helicity for all simulations. All combinations of $E$ and $Pm$ follow similar behavior with increasing $Ra$ and larger values of helicity are observed for decreasing Ekman numbers, whereas $Pm$ tends to have only a weak influence on $\He$. In our DNS we non-dimensionalize the equations using the depth $H$ and speed $\nu/H$; adapting this scaling to the asymptotic theory gives $|\ub'| = O(E^{-1/3})$ and $|\Zb'| = O(E^{-2/3})$ so that $|\He | = O(E^{-1})$ as $E \rightarrow 0$. Fig.~\ref{F:He}(b) shows the asymptotically rescaled rms helicity versus $\Rat$; the collapse of the data shows that these simulations are in a quasi-geostrophic dynamical state \citep{mC18}.

\begin{figure*}

\begin{minipage}{0.4\textwidth}
\centering
\subfloat[]{\label{main:a}\includegraphics[width=1\textwidth]{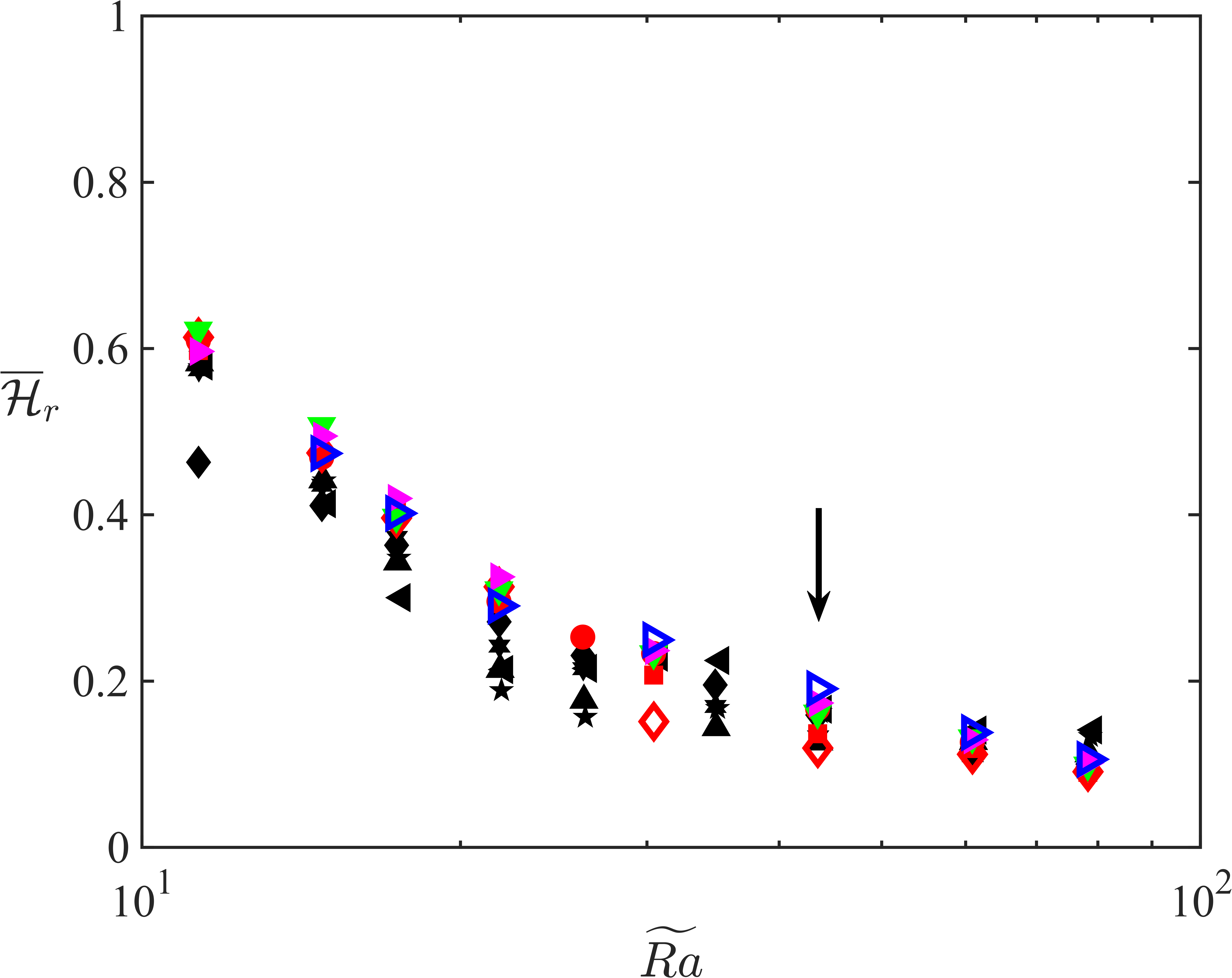}}
\end{minipage}%
\begin{minipage}{0.6\textwidth}
\centering
\subfloat[]{\label{main:b}\includegraphics[width=0.35\textwidth]{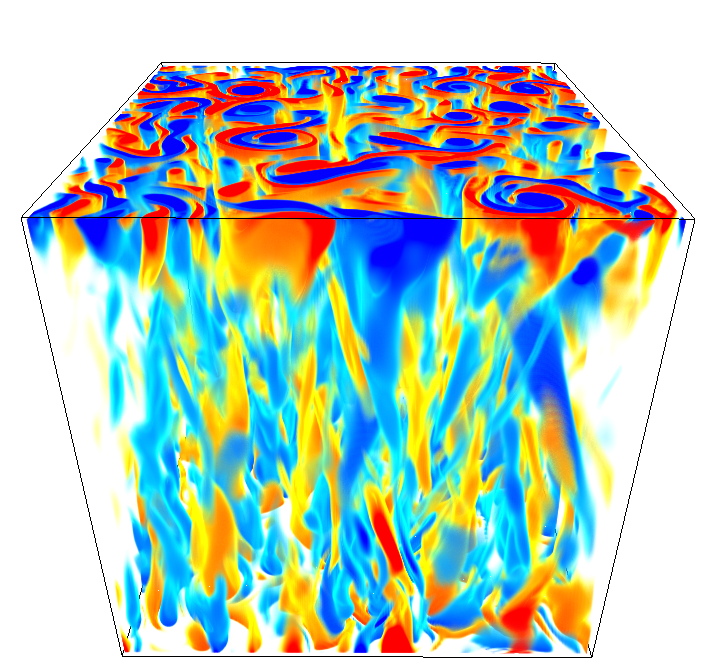}}
\subfloat[]{\label{main:b}\includegraphics[width=0.35\textwidth]{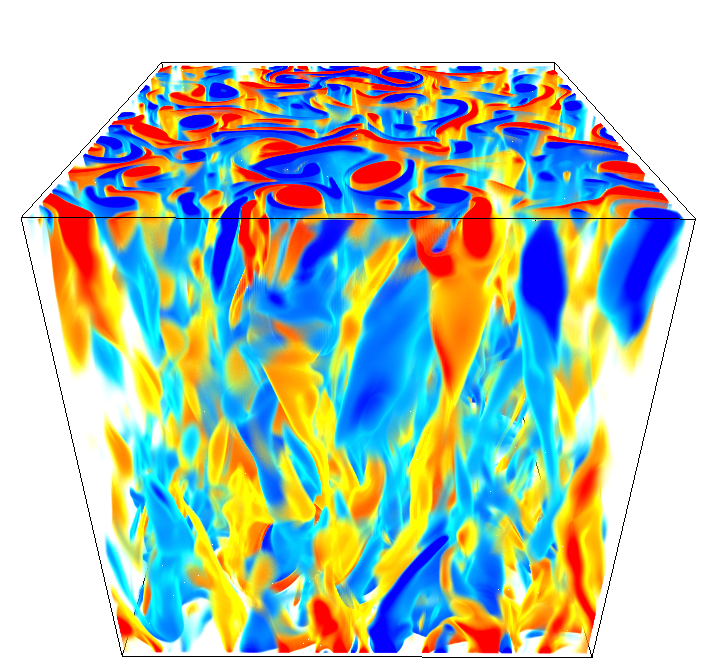}} \\
\subfloat[]{\label{main:b}\includegraphics[width=0.35\textwidth]{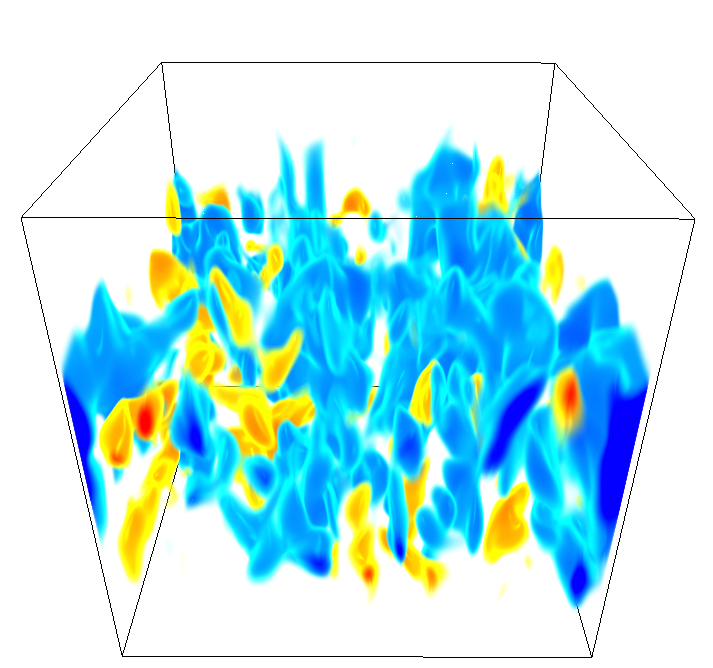}}
\subfloat[]{\label{main:b}\includegraphics[width=0.35\textwidth]{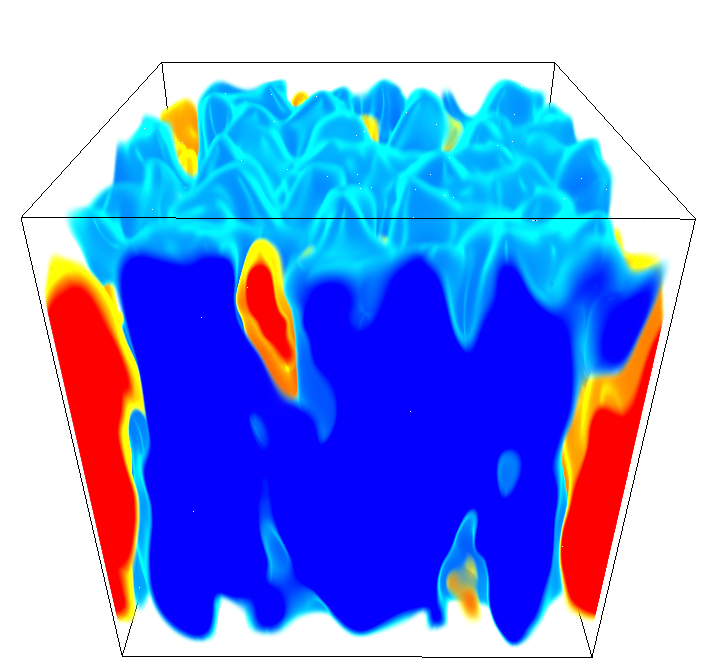}}
\end{minipage}\par\medskip

\caption{ (a) Relative (rms) helicity for all simulations versus $\Rat$; the arrow shows the parameter space location for the visualizations shown in (b)-(e). (b,d) Visualizations of (b) vertical vorticity and (d) $x$-component of the magnetic field for $E = 10^{-7}$, $Pm=0.3$ and $\Rat \approx 44$. (c,e) Visualizations of (c) vertical vorticity and (e) $x$-component of the magnetic field for $E = 10^{-8}$, $Pm=0.05$ and $\Rat \approx 44$. The symbols have the same meaning as in Fig.~\ref{F:He}. 
\label{F:He_rel}}
\end{figure*}

It is further helpful to define the relative helicity $\He_r = \He/ ( u \, \zeta  ) $, where $u$ and $\zeta$ are volumetric rms values, such that maximally helical flows are characterized by $| \He_r | = O(1)$. In agreement with previous studies, Fig.~\ref{F:He_rel}(a) shows that $| \He_r |$ becomes small as $\Rat$ increases for all of the simulations, yet remains finite \citep[e.g.][]{sST10,kS12,mC16b}. These results suggest that natural dynamos, for which $\Rat \gg 1$, are also characterized by small values of the relative helicity. However, reducing the magnetic Prandtl number shows that despite the small relative helicity, large scale dynamo action remains possible, as predicted by theory. Figs.~\ref{F:He_rel}(b)-(e) compare flow visualizations for (b,d) $E = 10^{-7}$ and $Pm=0.3$ with (c,e) $E = 10^{-8}$ and $Pm=0.05$. As indicated by the black arrow in Fig.~\ref{F:He_rel}(a), both cases are characterized by similar values of $\He_r$. The renderings of the vertical component of the vorticity shown in (b) and (c) provide visual evidence that both simulations are characterized by dynamically similar turbulent flows. Whereas the large scale Reynolds numbers are $Re \approx 4.5 \times 10^3$ and $Re \approx 1.1 \times 10^4$ for $E = 10^{-7}$ and $E = 10^{-8}$, respectively, the corresponding small scale Reynolds numbers for both cases are $\Ret \approx 21$ and $\Ret \approx 24$. The corresponding $x$-component of the magnetic field vector is shown for the two cases in Figs.~\ref{F:He_rel}(d) and (e) where we find a more obvious coherent large scale component for the $E = 10^{-8}$ ($Pm=0.05$) case.

 \begin{figure*}
 \begin{center}
                  \subfloat[]{
      \includegraphics[width=0.45\textwidth]{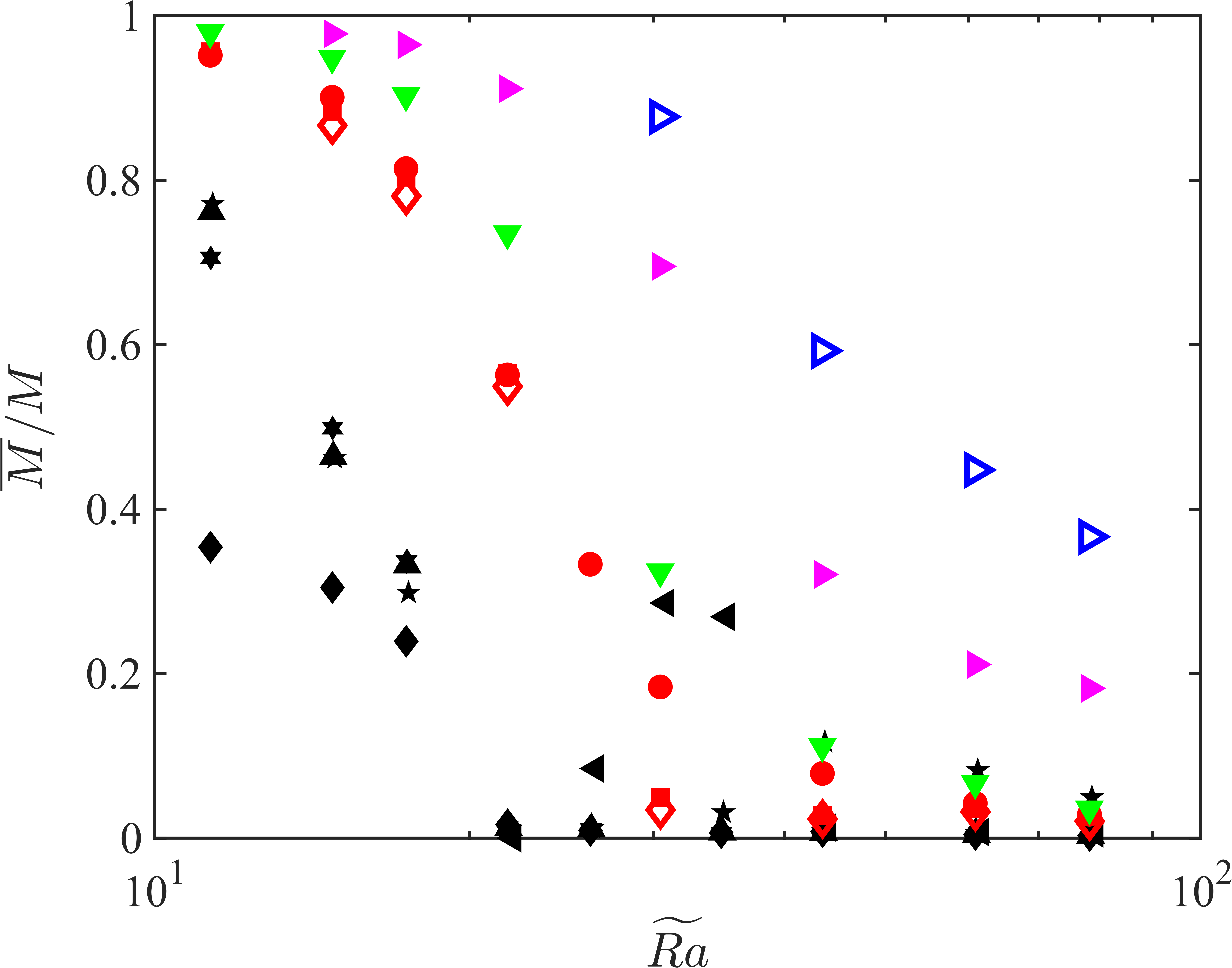} } 
      \quad
                 \subfloat[]{
      \includegraphics[width=0.45\textwidth]{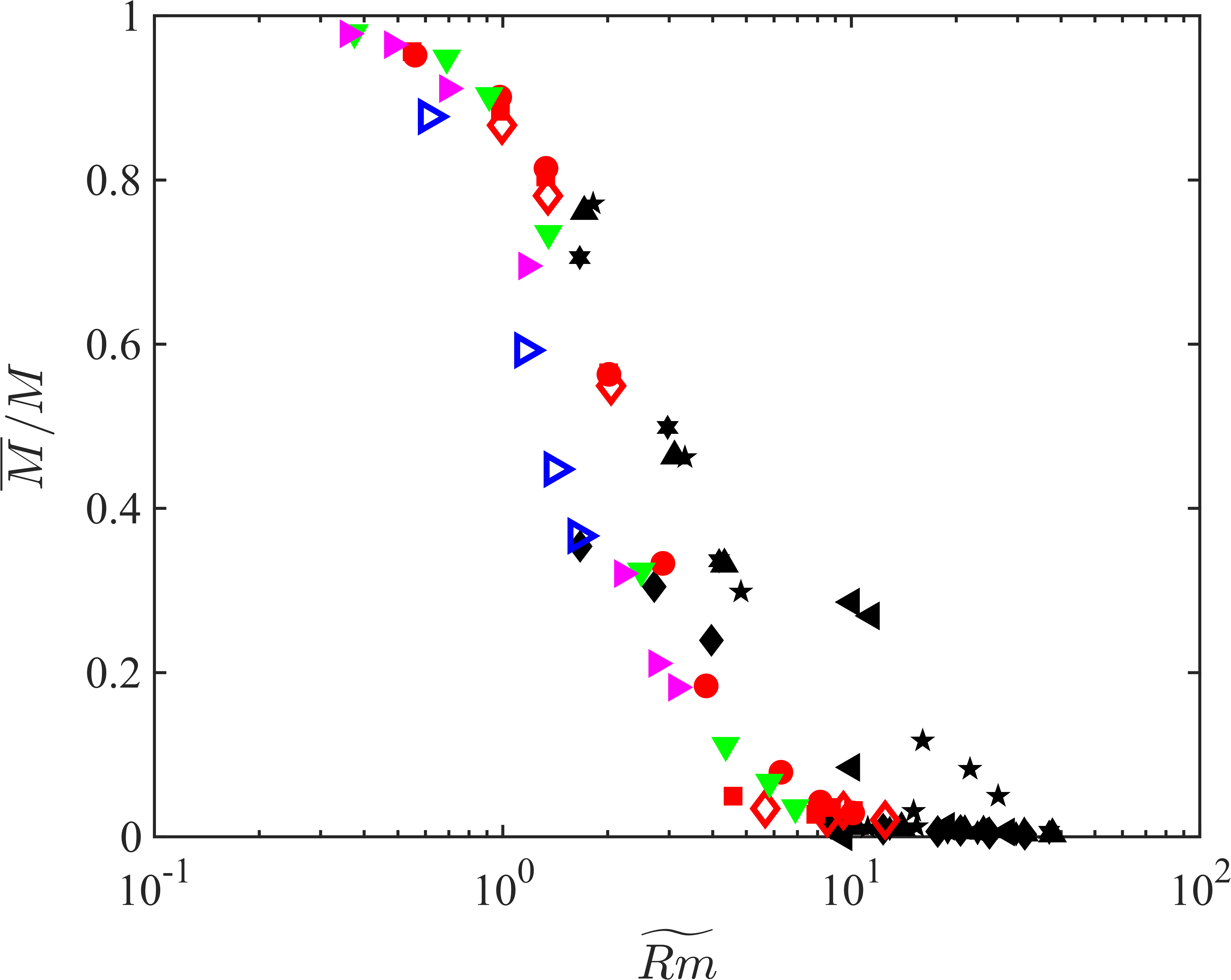} }         
 \end{center}
\caption{ Fraction of the mean magnetic energy to the total magnetic energy, $\overline{M}/M$, in all simulations: 
(a) $\overline{M}/M$ versus $\Rat$;
(b) $\overline{M}/M$ versus $\Rmt$.
The symbols have the same meaning as in Fig.~\ref{F:He}.
\label{F:Emfraction}}
\end{figure*} 

The relative size of the large scale magnetic field can be quantified by computing $\overline{M}/M$, where $\overline{M}$ is the magnetic energy of the mean magnetic field and $M$ is the total magnetic energy. Fig.~\ref{F:Emfraction}(a) shows this mean energy fraction versus $\Rat$. We find that smaller values of $Pm$ typically yield larger values of $\overline{M}/M$ for a given value of $\Rat$. For a fixed value of $Pm$ the mean energy fraction decreases with increasing $\Rat$. Conversely, for  a fixed value of $\Rat$, decreasing $Pm$ typically yields larger values of $\overline{M}/M$. Fig.~\ref{F:Emfraction}(b) shows that a collapse of the data occurs when $\overline{M}/M$ is plotted versus $\Rmt$. In particular, we find energetically robust large scale dynamos, as characterized by $\overline{M}/M  \rightarrow 1$, only when $\Rmt \lesssim O(1)$.

We note that the data shown in Fig.~\ref{F:Emfraction}(b) also demonstrates that dynamo action is achieved for smaller values of $\Rmt$ as both $E$ and $Pm$ are reduced. This reduction in the value of $\Rmt$ needed for dynamo action as the parameter values are made more extreme is because these dynamos are intrinsically multiscale and anisotropic; convective motions take place over the depth $H$ of the system, yet the small $O(H E^{1/3})$ horizontal length scale of the convection is crucial to these dynamics. Although $\Rmt$ is becoming smaller as both $E$ and $Pm$ are reduced, the large scale magnetic Reynolds number scales as $Rm = O(E^{-1/6})$, and therefore becomes large even as $\Rmt = O(E^{1/6})$ becomes small as $E \rightarrow 0$. 

\begin{figure*}
 \begin{center}
      \includegraphics[width=0.45\textwidth]{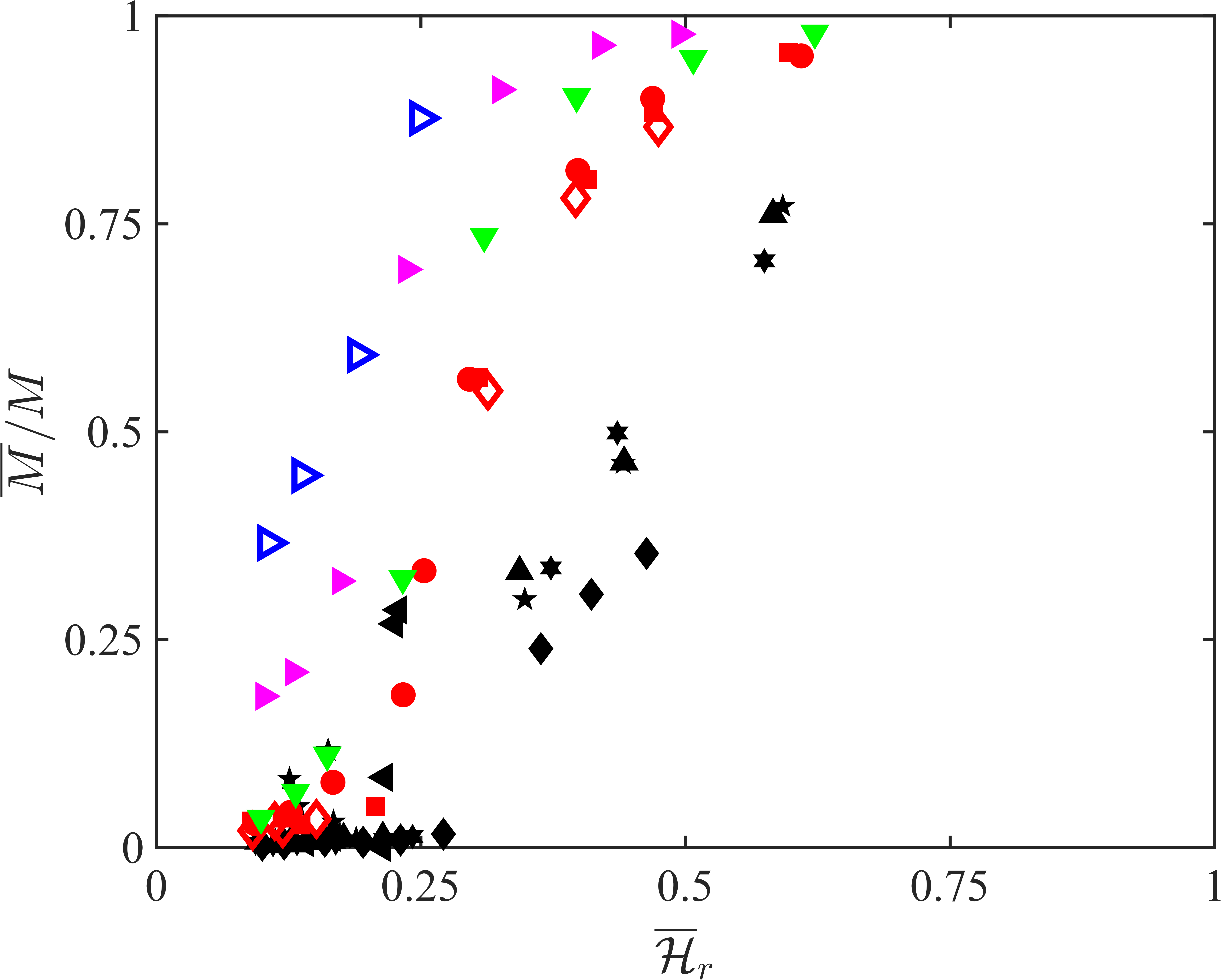} 
 \end{center}
\caption{ The fraction of the mean magnetic energy to the total magnetic energy $\overline{M}/M$ versus  the rms relative helicity $\He_r$. The symbols have the same meaning as in Fig.~\ref{F:He}.
\label{F:Emfraction_rHe}}
\end{figure*} 

The relationship between the mean energy fraction and relative helicity is shown in Fig.~\ref{F:Emfraction_rHe}. We find that robust large scale dynamo action is achieved at smaller values of the relative helicity as both $E$ and $Pm$ are reduced. These findings show that small relative helicity by itself does not imply that large scale dynamo action is not achievable, and that a deficit of helicity can be offset by enhancing the influence of magnetic diffusion on the small convective length scale. A possible explanation for these observations is that the stretching-diffusion balance present in the small scale induction equation prevents small scale magnetic field from cascading to smaller length scales. We note that simulations of non-rotating helical turbulence have found a similar effect in which large scale dynamos can be maintained with decreasing helicity so long as the forcing wavenumber is increased \citep{jG12}.

In comparison to previous studies of rotating dynamos in the plane parallel geometry, the present investigation extends the parameter space to smaller Ekman numbers and smaller magnetic Prandtl numbers. Refs.~\citep{cG15,cG17} find that so-called large scale vortices (LSVs), generated by the inverse kinetic energy cascade mechanism, can play an important role in sustaining large scale magnetic field. However, none of the cases reported in Refs.~\citep{cG15,cG17} show evidence of what we refer to as the energetically robust large scale dynamo regime in which $\overline{M}/M  \approx 1$. This finding is consistent with the trends observed in our simulations. Moreover, no general conclusion on the importance of LSVs in our simulations could be reached; whereas some cases show relatively strong LSVs, in many of the cases reported they are not energetically dominant, similar to the findings of Ref.~\citep{pB18}. Ref.~\citep{aT12} found transitions in dynamo behavior for $Pm \ge 1$ when $\Rmt \gtrsim 13$, though we do not observe a similar behavior that may be due to the smaller Ekman and Rossby numbers employed here. Ref.~\citep{pK09} finds large scale dynamos for sufficiently small Ekman numbers and $Pm \ge 1$, though energetically robust large scale dynamos are not observed.
 
The combination of the plane parallel geometry and the Boussinesq approximation has enabled a systematic exploration of parameter space in which $E \ll 1$ and $Pm < 1$. Sphericity and compressibility are both important physical effects in natural systems such as stars and planets. However, due to their global nature, spherical simulations require enormous resolution requirements as the Ekman number reduced \citep[e.g.][]{nS17}. Small Mach number compressible convection simulations are regularly used \citep{pK09,bF13,pB18}, but the increased computational cost relative to Boussinesq simulations nevertheless prevents such studies from accessing values of $E$ and $Pm$ comparable to those studied here. We note that neither compressibility \citep{mC15} nor sphericity \citep{nS17} influence the leading order balances on the small convective length scales, suggesting that the distinguished limit identified in Ref.~[10] and studied here may still play the key role in controlling the relative size of the large scale magnetic field.


The authors gratefully acknowledge funding from the National Science Foundation (NSF) through grants EAR-1620649 (MY and MAC), EAR-1945270 (MAC) and SPG-1743852 (MAC). This work used the Extreme Science and Engineering Discovery Environment (XSEDE) Stampede2 supercomputer at the Texas Advanced Computing Center (TACC) through allocation PHY180013. Additional computing resources were provided on the RMACC Summit supercomputer, which is supported by the NSF (awards ACI-1532235 and ACI-1532236), the University of Colorado Boulder, and Colorado State University. Flow visualization was performed with VAPOR \citep{sL19}.

\newcommand{\jfm}{J. Fluid Mech.~}\newcommand{\mnras}{Mon. Not. Roy. Astron.
  Soc. }\newcommand{\jgr}{J. Geophys. Res.~}\newcommand{\araa}{Annu. Rev.
  Astron. Astrophys.~}\newcommand{\icarus}{Icarus }\newcommand{\aap}{Astron.
  Astrophys.~}\newcommand{\physscr}{Phys. Scripta }\newcommand{\ssr}{Space Sci.
  Rev.~}\newcommand{\pnas}{Proc. Nat. Acad. Sci.~}\newcommand{\ncom}{Nat.
  Comm.~}\newcommand{\njp}{New J. Phys.~}\newcommand{\prf}{Phys. Rev. Fluids
  }\newcommand{\prr}{Phys. Rev. Res. }\newcommand{\pepi}{Phys. Earth Planet.
  Int.~}\newcommand{\gji}{Geophys. J. Int.~}

\end{document}